\documentclass{article}
\usepackage[T1]{fontenc} 
\usepackage[utf8]{inputenc} 
\usepackage{ismir,amsmath,cite,url}
\usepackage{graphicx}
\usepackage{color}
\usepackage{mathtools}
\usepackage{multirow}
\usepackage{pbox}
\usepackage{caption}
\usepackage{subcaption}
\usepackage{lineno}
\usepackage[ruled,noend]{algorithm2e}
\usepackage{soul}
\usepackage{hyphenat}
\title{Modeling the rhythm from lyrics for melody generation of pop song}

\multauthor
 {Daiyu Zhang \hspace{0.3cm} Ju-Chiang Wang \hspace{0.3cm} Katerina Kosta \hspace{0.3cm} Jordan B. L. Smith \hspace{0.3cm} Shicen Zhou}
{
ByteDance \\
{\tt\small \{daiyu.zhang, ju-chiang.wang, katerina.kosta, jordan.smith, zhoushicen\}@bytedance.com}
}

\def\authorname{D. Zhang, J.-C. Wang, K. Kosta, J. B. L. Smith, and S. Zhou}

\begin{document}

\maketitle
\begin{abstract}
Creating a pop song melody according to pre-written lyrics is a typical practice for composers. A computational model of how lyrics are set as melodies is important for automatic composition systems, but an end-to-end lyric-to-melody model would require enormous amounts of paired training data. To mitigate the data constraints, we adopt a two-stage approach, dividing the task into lyric-to-rhythm and rhythm-to-melody modules. However, the lyric-to-rhythm task is still challenging due to its multimodality. In this paper, we propose a novel lyric-to-rhythm framework that includes part-of-speech tags to achieve better text-setting, and a Transformer architecture designed to model long-term syllable-to-note associations. For the rhythm-to-melody task, we adapt a proven chord-conditioned melody Transformer, which has achieved state-of-the-art results. Experiments for Chinese lyric-to-melody generation show that the proposed framework is able to model key characteristics of rhythm and pitch distributions in the dataset, and in a subjective evaluation, the melodies generated by our system were rated as similar to or better than those of a state-of-the-art alternative.

\end{abstract}
\section{Introduction}
\label{sec:introduction}


Setting lyrics to a melody is a common but complex task for a composer.
The form, articulation, meter, and symmetry of expression in lyrics can inspire, or set constraints on, the melodic arrangement.
Given the importance of melody, it is unsurprising that the decades-long history of Music Metacreation systems includes countless melody-creation systems (see~\cite{pasquier2017introduction} for a review).
However, less attention has been paid to the lyric-to-melody generation task (i.e., generating a melody for given input lyrics).
The task is challenging for many reasons, including but not limited to: the need to handle the prosody of the text correctly (e.g., one should avoid setting an unstressed word like `the' on a stressed note in the melody); the need to reflect the structure of the lyrics in the melody; and the need to create a good melody to begin with.

With the rapid growth of deep learning tools, this task has gained more attention, and there are many recent examples of lyric-to-melody creation systems, most using an end-to-end approach \cite{yu2021conditional,sheng2021songmass,bao2019neural,lee2019icomposer}.
Modeling the relationship between lyric syllables and musical notes is a complex, cross-modal task,
but it is hoped that
we can succeed with a large amount of paired examples (i.e., lyrics aligned to their corresponding melodies).
However, acquiring such data is expensive, and using unsupervised learning has shown limited performance gains \cite{sheng2021songmass}.
All the systems mentioned here are trained on fewer than 200,000 examples of song lyrics; by contrast, the text-to-image system DALL-E has 12 billion parameters and involved hundreds of millions of paired text-image training data~\cite{ramesh2021zeroshot}.

\begin{figure}
  \centering
  \includegraphics[width=\columnwidth]{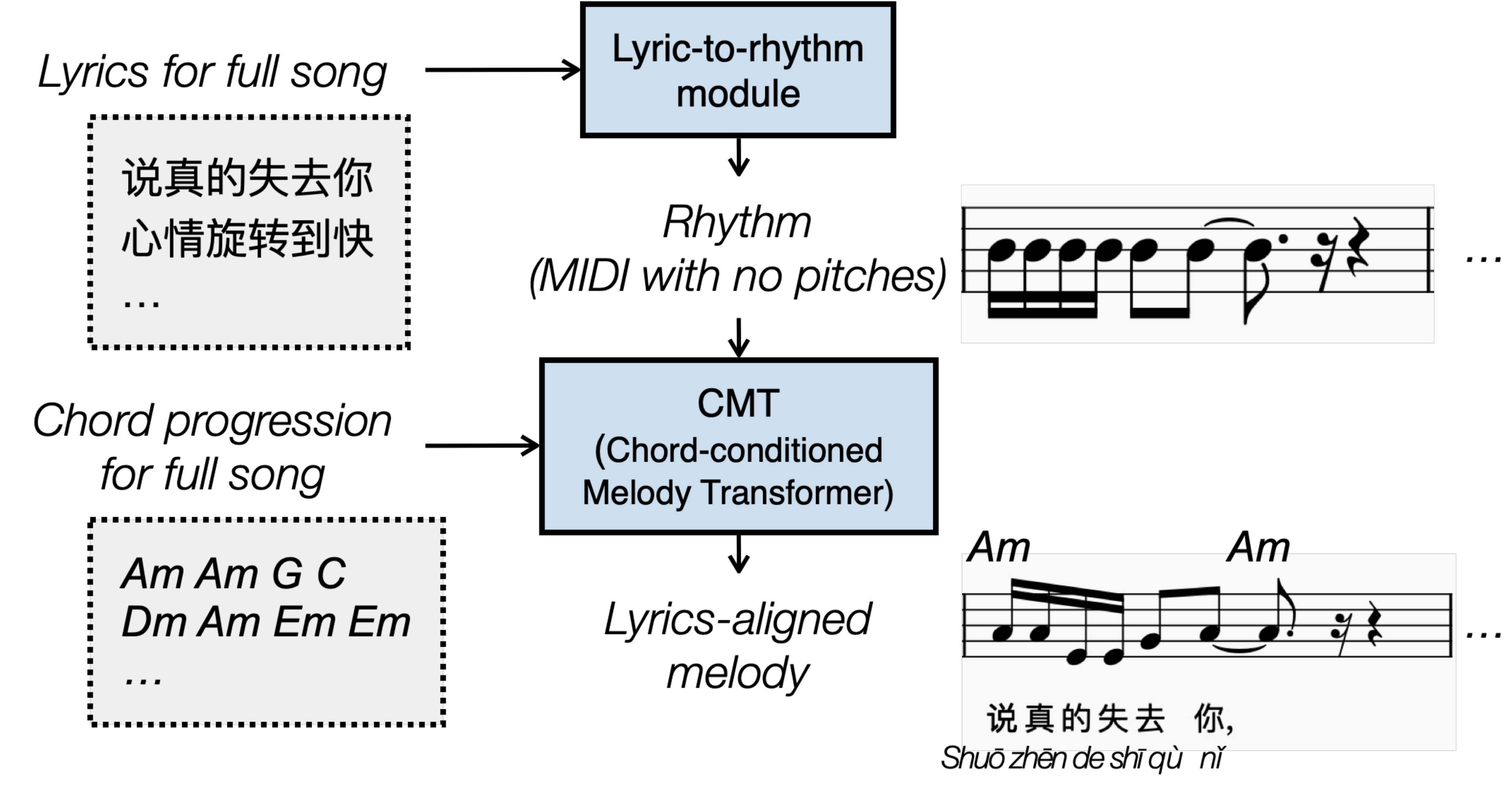}
  \caption{Diagram of the proposed system.}
  \label{fig:system}
\end{figure}

One alternative, suggested in \cite{ju2021telemelody}, is to pick an intermediate representation and adopt a two-stage approach: one model to convert lyrics to the chosen representation, and a second to convert that to a melody.
The motivation is that there is sufficient data to train each model separately, without the paired lyrics-melody data required by the end-to-end approach.

We choose `rhythm' as the intermediate step because, if we disregard melismas and expressive singing techniques, we can assume there is a one-to-one correspondence between syllables and onsets, and between onsets and melody pitches.
Also, there is plenty of data to model each step: first, from karaoke-style scrolling lyrics data, we can obtain an alignment between syllables in lyrics and note onsets in music, and thus note durations and metrical positions, too. Second, there are multiple public datasets from which to learn to assign pitches for each note given their duration.
Our goal is then to solve two sub-tasks, namely lyric-to-rhythm and rhythm-to-melody, with an assumption that the rhythm generation process is independent of the pitch generation one \cite{ju2021telemelody}.

There are many recent melody generation models~\cite{chen2020music, dai2021controllable, chen2020continuous, tan2019chordal},
but lyric-to-rhythm modeling is rarely attempted.
In this paper, we introduce a novel framework for converting lyrics to rhythms using an encoder-decoder Transformer architecture \cite{vaswani2017attention}.
The proposed system is outlined in Fig.~\ref{fig:system}: given an input set of lyrics, a lyric-to-rhythm module assigns onset times and durations for each syllable. This rhythm, along with a user-provided chord progression, is fed into a Chord-conditioned Melody Transformer (CMT) \cite{choi2021chord}, a state-of-the-art melody generation system, to predict the pitch for each note.
The details of the lyric-to-rhythm module and the CMT are provided in Sections~\ref{subsec:lyric2rhythm} and~\ref{subsec:cmt}, respectively.

\section{Background}
\label{sec:related}
Lyrics and melody are not arbitrarily combined; common sense suggests and prior analysis~\cite{nichols2009relationships} indicates that patterns in lyrics and melodies are related and can be modeled, in part, with features of the melody (e.g., note duration) and lyrics (e.g., syllable stress).
One of the earliest lyric-to-melody systems was designed to handle Japanese prosody~\cite{fukayama2010automatic}: first, the input text was segmented into phrases; next,
a set of pre-composed rhythms was searched for one that fit the syllable count and matched the accent pattern of the text; 
finally, pitches were assigned using dynamic programming to optimise the interval directions with the natural prosody of the words. An earlier lyric-to-rhythm system also leveraged a dataset of pre-composed rhythms that were scored based on their match to the input syllable-stress and word-rarity patterns~\cite{nichols2009lyric}. Although our system has little in common with these works, we do share the use of rhythm as an intermediate representation.

Algorithms for automatic music generation are a subset of Music Metacreation systems~\cite{pasquier2017introduction}, which have been present in Western music in many forms, including being used for the creation of standalone pieces and, either offline or in real-time, as part of the human composition process. 
With the help of machine learning and deep learning architectures, many such systems have shown to be capable of generating a plausible outcomes that match the musical characteristics of given datasets. 
Supervised generative models aim to learn a representation of the underlying characteristics of a training set distribution. 
Depending on the model, this representation can be either explicitly depicted or implicitly used to generate samples from the learned distribution~\cite{goodfellow}. 

Some systems aim to generate a part of a musical piece with the aid of another given part (including melody-to-lyrics creation~\cite{ma2021ailyricist}, the inverse of the task we consider). Conditioning the choice of parameters in a generative model on data from other modalities, such as a bass line or a structure, can yield controllable generation systems~\cite{briot2020deep}[p.82-83]. For the case of using chords to condition melody generation, a recent system adjusting a general adversarial network architecture has been presented in~\cite{midinet} with the option of generating melody lines over a given accompaniment. The Chord-conditioned Melody Transformer (CMT) \cite{choi2021chord} is the most recent effort in this area; we adapt much of the design of this system, extending it to accept both lyrics and chords as input. Details of this system, and how we adapt it, follow in Section~\ref{sec:method}.

\section{Methodology}
\label{sec:method}

\subsection{System Overview}
\begin{figure}
  \centering
    \includegraphics[width=\columnwidth]{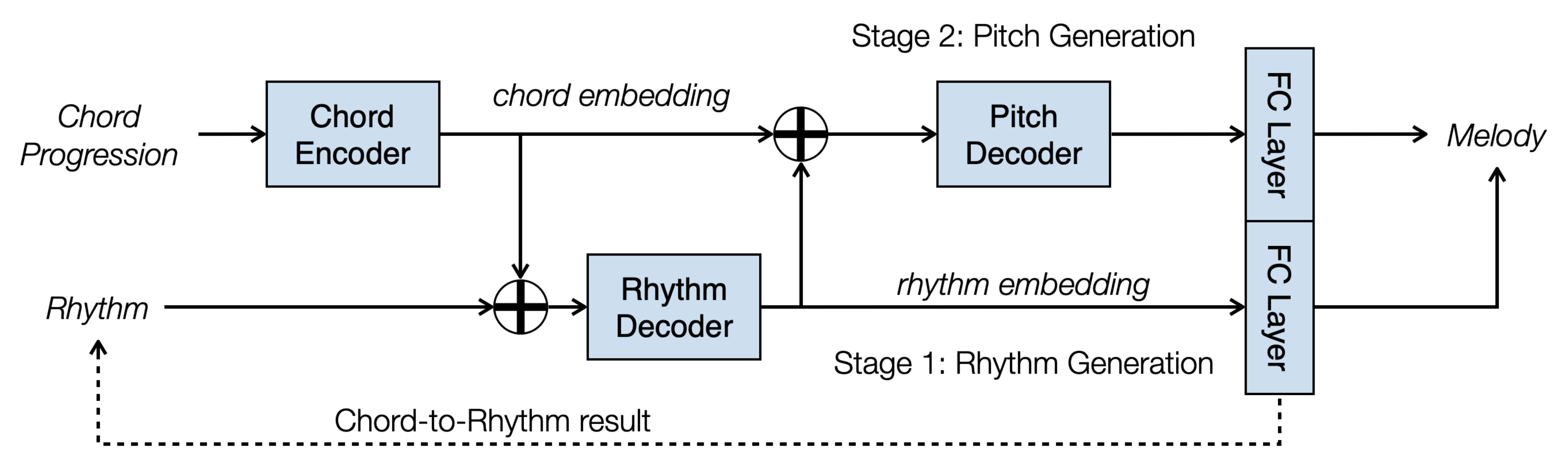}
  \caption{A two-stage structure of CMT, where $\oplus$ represents concatenation. Stage 1: chord-to-rhythm. Stage 2: rhythm+chord-to-pitch based on the result of Stage 1.}
  \label{fig:cmt}
\end{figure}

Our system design is motivated by the Chord-conditioned Melody Transformer (CMT) \cite{choi2021chord}. The authors of CMT proposed a two-stage system, assuming a hierarchy that the process of generating melodies is two-phase, as depicted in Fig.~\ref{fig:cmt}: \emph{Stage 1}, generating the rhythm of notes from chord progressions; \emph{Stage 2}, generating the pitch for each note depending on the chord progressions and generated rhythm. Our proposed system augments CMT by replacing chord-to-rhythm (i.e., Stage 1) with a novel \emph{lyric-to-rhythm module}. As a result, users can input the lyrics and chord progression of a full song in our system (see Fig. \ref{fig:system}). Then, the lyric-to-rhythm module generates the MIDI (with empty pitches). Second, CMT processes the MIDI and chord progression to generate the melody.
As a result, the rhythm is generated with a global view of the lyrics, while the melody is generated with a causal view of the rhythm and chords.
%

In the following subsections, we will first review CMT and explain the difficulties of modifying it to handle the lyric-to-melody task in Section \ref{subsec:cmt}. Then, we will detail our solution in Section \ref{subsec:lyric2rhythm}.

\subsection{Chord-Conditioned Melody Transformer (CMT)}
\label{subsec:cmt}

CMT adopts a pianoroll-like representation ~\cite{yang2019deep,choi2021chord} that includes chord, rhythm, and pitch (CRP) information. It splits the timeline into semiquaver-length frames (1/4 of a beat), each described by three vectors: a 12-dimensional binary \textit{chord} vector (pitch classes in the chord get a 1); a 3-dimensional one-hot \textit{rhythm} vector (onset, hold state, rest state); and a 22-dimensional one-hot melodic \textit{pitch} vector (for this part we restrict MIDI pitches to between 48 and 67, plus a hold state and a rest state, giving a total dimension of 22).
Please refer to \cite{choi2021chord}[Fig. 1] for an illustration. 


CMT contains three main modules: \emph{Chord Encoder (CE)}, a bidirectional LSTM \cite{graves2005bidirectional}; \emph{Rhythm Decoder (RD)}, a stack of self-attention blocks; and \emph{Pitch Decoder (PD)}, another stack of self-attention blocks. In Stage 1, given an input chord progression, the chord embedding encoded by CE is autoregressively sent into RD to output the rhythm embedding, followed by a fully-connected layer (``FC Layer'' in Fig.~\ref{fig:cmt}) to predict the sequence of rhythm vectors for the entire song. In Stage 2, the concatenation of the chord and rhythm embeddings is autoregressively fed into PD, followed by a fully-connected layer to predict the sequence of pitch vectors. Finally, rhythm and pitch vectors are combined and converted to the melody. 


However, to leverage CMT for the lyric-to-melody task, we face three problems. (1) \emph{Multimodality}: CMT was designed to take the input of a chord progression to generate the melody. However, it is non-trivial to directly add a lyric encoder for lyrics input, as lyrics are more complicated sequential data than chords. (2) \emph{Representation}: CMT uses a pianoroll-like (i.e. CRP) representation to encode melody, where the time axis is evenly scaled (e.g., 1/16 beat), so a note may require multiple tokens to carry the duration. This makes it difficult to create a one-to-one mapping that ties a syllable (or character) to a single note token. (3) \emph{Constraint on length}: in CMT, the CE generates the melody on a segment-to-segment basis (e.g., 8 bars at a time) without exploiting the global context of a full-song chord progression. However, we believe the structural information carried in the input lyrics is crucial to determine the repetitive pattern for the output melody.
The next subsection details how we address these problems: (1) is addressed with a POS tagger that compactly encodes useful lyrics information; and (2) and (3) are addressed by adapting a Compound Word representation.

\begin{figure}
  \centering
  \includegraphics[width=\columnwidth]{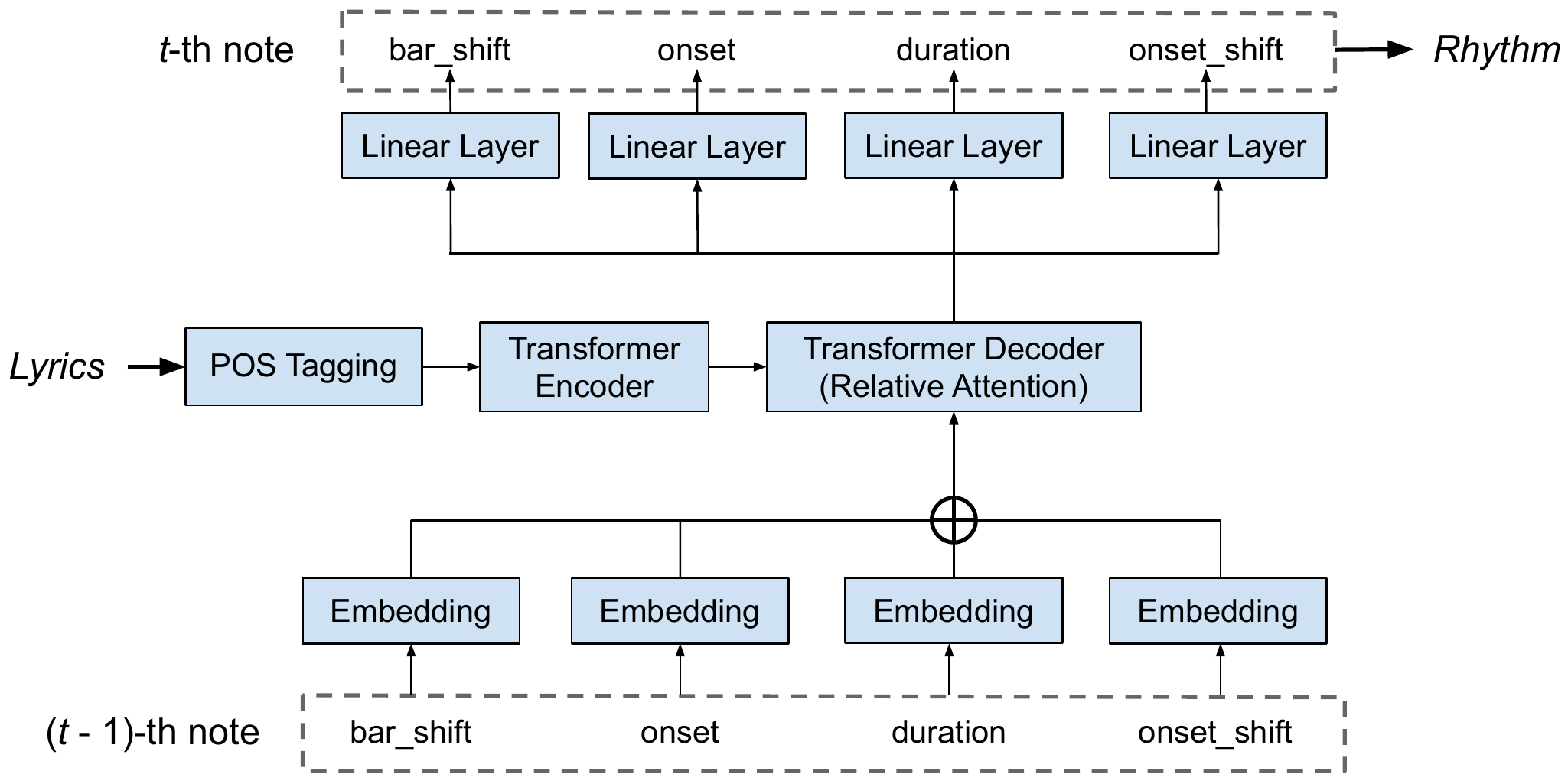}
  \caption{The proposed lyric-to-rhythm framework.}
  \label{fig:l2r}
\end{figure}

\subsection{Lyric-to-Rhythm Framework}
\label{subsec:lyric2rhythm}

Fig.~\ref{fig:l2r} shows our lyric-to-rhythm framework, which is analogous to a language translation task: i.e., an input sequence of lyrics is translated into an output sequence of notes. In this work, we assume that each syllable (or character) is mapped onto one note as a simplification; handling melismas remains a future challenge.
To this end, we adapt an encoder-decoder Transformer architecture \cite{vaswani2017attention}. To enhance the repetitive coherence modeling in note sequences, we incorporate relative self-attention \cite{shaw2018self,huang2018music}.

To extract the features of lyrics, we employ \emph{part-of-speech (POS) tagging} with a Transformer encoder.
Following prior works \cite{zeng2021musicbert,chou2021midibert}, we characterize the rhythmic features of a note with a tuple of \texttt{\footnotesize{(bar\_shift, position, duration, onset\_shift)}}, and model the sequence of tuples using the \emph{Compound Word (CP)} Transformer decoder \cite{hsiao2021compound}. The lyric-to-rhythm module generates the $t$-th note based on the full context of lyrics and the previously generated notes (from the first to ($t$-1)-th) in an auto-regressive manner, with the future notes being masked. We describe the POS tag representation and CP Transformer in the next subsections.

\subsubsection{Part-of-Speech (POS) Tagging}

In natural language processing, POS tagging refers to the process of labeling every word in a text with its part of speech. The taxonomy of POS tags varies by language, but commonly includes `noun,' `verb,' `adjective,' `adverb,' and others. POS tags can augment the text information by indicating the structure of sentences \cite{sun2011improved}, and thus plays an important role in tokenizing the input words in conventional text-to-speech (TTS) systems \cite{jurafsky2008speech,zen2007hmm}.

POS are word-level descriptors, but we want syllable-level descriptors in order to align the lyrics with the rhythm.
(When dealing with Chinese lyrics, we can also say `character-level' since each Chinese character is one syllable.)
Thus, we combine each POS tag with the syllable index to create a POS 'token': e.g., the input English sentence ``Why not tell someone,'' would result in: [`adverb-0', `adverb-0', `verb-0', `noun-0', `noun-1'], where the two syllables in ``someone'' are represented by [`noun-0', `noun-1'].

\subsubsection{Compound Word Transformer}

\begin{table}
\small \centering
 \begin{tabular}{l|l|l}
Token Name & Vocab. & Description \\
\hline
\texttt{\footnotesize{bar\_shift}} & 0, 1, 2 & Time shift in bar to current bar  \\
\texttt{\footnotesize{onset}} & 0 -- 15 & Onset in 1/4 beat in current bar  \\
\texttt{\footnotesize{duration}} & 0 -- 31 & Duration in 1/4 beat  \\
\texttt{\footnotesize{onset\_shift}} & 0 -- 15 & \pbox{10cm}{Time shift in 1/4 beat to previous \\ note's onset} \\
 \end{tabular}
\caption{Rhythmic features for a note.}
\label{tab:compund_word}
\end{table}

In contrast to the CP proposed in \cite{hsiao2021compound}, we do not distinguish between \emph{note}- and \emph{metre}-related events.
Instead, we include four tokens in every compound word (see Table \ref{tab:compund_word}), so that we can have one set of tokens per syllable.
From karaoke scrolling lyrics data, we can obtain the onset and duration of each syllable, and by tracking the downbeats, we can obtain the metric position in the bar.
Following \cite{hsiao2021compound}, each of the four tokens is converted into an embedding, and then the embeddings are concatenated before being sent to the Transformer decoder. Each of the four output embeddings is linearly projected to predict the value for the associated token of the $t$-th note. 

Once all the notes are ready, we convert them to MIDI (with unspecified pitch) with the following steps:
\begin{enumerate}
    \setlength{\itemsep}{-4pt}%
    \item Place an empty note at bar 0 and position 0. 
    \item Determine the onset by shifting (\texttt{\footnotesize{bar\_shift}}$\times16 +$  \texttt{\footnotesize{onset\_shift}}) units from the previous onset.
    \item Set the duration by $min($\texttt{\footnotesize{duration}}, next note's \texttt{\footnotesize{onset\_shift}}$)$.
    \item Repeat 2 and 3 until all the notes are processed.
\end{enumerate}
We note that \texttt{\footnotesize{onset}} is not used for generating MIDIs. Instead, we use the position shifted to determine the onset so that notes are placed in an incremental order. Nevertheless, we suspect that \texttt{\footnotesize{onset}} can help regularization in training.
Using the CP representation addresses the ``constraint on length'' issue mentioned in Section \ref{subsec:cmt}, as
it permits a more compact sequence of tokens that can model a longer duration, such as a full song.

\section{System Configuration}
\label{sec:exp_set}

This section describes how we trained the lyric-to-rhythm and rhythm-to-melody models.
For each model we explain what data were used and how they were collected.
We focus on Chinese pop songs to validate our system, but the framework could be adaptable to other languages since parts of speech and syllables are broadly useful concepts.

\subsection{Lyric-to-Rhythm Model}
\label{subsec:l2r}

We collected data for 45K Chinese pop songs using a similar pipeline as~\cite{xue2021deeprapper} and~\cite{ju2021telemelody}[Appendix A].
That is, we crawled online to obtain paired lyrics and audio, with timestamps indicating the onset of each line of the lyrics.
Then, for each song, we performed the following steps:
(1) isolate the vocal audio using source separation;
(2) convert lyrics to phoneme sequences;
(3) estimate the phoneme onset timestamps using forced phoneme alignment;
and (4) estimate the time signature and beat and downbeat times.
From the phoneme and beat data, we can derive the syllable onsets and thus the bar-shift, onset, duration and onset-shift attributes required by the model.
The steps were performed using in-house tools comparable to those used in~\cite{ju2021telemelody}: Spleeter~\cite{spleeter2020}, Phonemizer\cite{bernard2021phonemizer}, Montreal Forced Aligner~\cite{mcauliffe2017montreal}, and Madmom~\cite{madmom}, respectively.

We kept songs with a detected time signature of 4/4 (around 90\%), and quantized the timestamp of each syllable in quarter beats.
Errors in automatic lyrics alignment, in beat tracking, and in the detected time signature can all degrade the model quality, so we selected 330 songs to manually adjust the timestamps.
This subset was used to fine-tune the 
model.


For POS tagging, we adopted Jieba\footnote{\url{https://github.com/fxsjy/jieba}}, an open-source tool that supports 56 tags commonly used in Chinese.
Without POS tags, the vocabulary size for our dataset was 5,368 unique characters.
Reducing to POS and then adding the syllable index resulted in a vocabulary of 123 unique POS tokens.
In Sec.~\ref{subsec:subjective_results}, we will compare a model using this 123-dimensional POS vector to an ablated version of the system that encodes the raw characters index in a 5368-dimensional vector.

\begin{figure}
  \centering
  \includegraphics[width=\columnwidth]{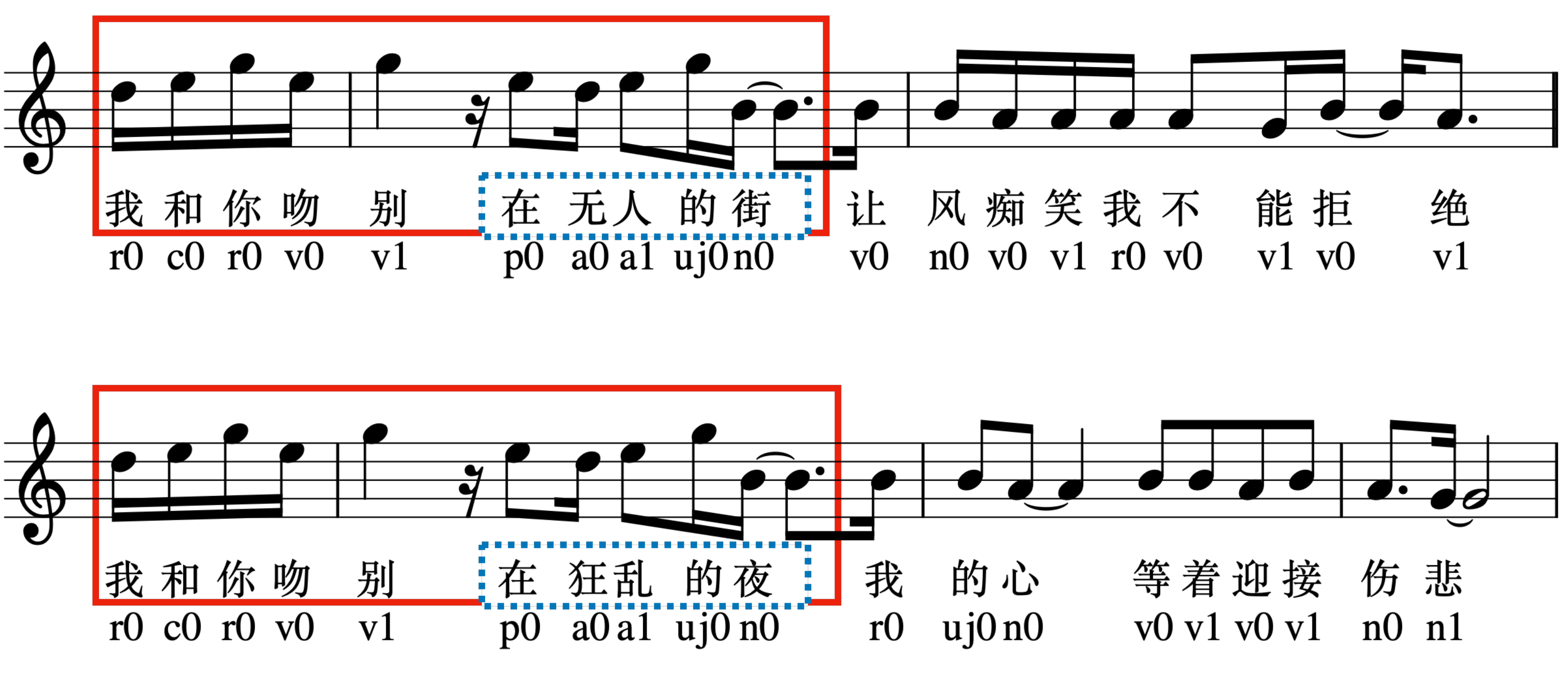} 
  \caption{The melody-lyrics-POS example in the chorus section of ``Goodbye Kiss''. POS abbreviation key:  \{`r': pronoun, `c': conjunction, `v': verb, `p': preposition, `a': adjective, `n': noun, `uj': auxiliary\}.}
  \label{fig:kissbye}
\end{figure}

In Chinese pop songs, symmetric expression of text structure is commonly reflected in melody repetition. Fig.~\ref{fig:kissbye} shows one example: the chorus melody of ``Goodbye Kiss''\footnote{\url{https://www.youtube.com/watch?v=bJRkEmrkIO4}} by Jacky Cheung.
The two phrases outlined in solid boxes are identical in melody, and nearly identical in text; but even where the lyrics are different (in the dotted boxes), they have the same POS tags.
With the POS tagging representation, we believe the lyric-to-rhythm model can learn to generate similar rhythms for two text phrases if they have a common structure.

We use the following parameters for the encoder-decoder Transformer:
the input length is 1000; the numbers of heads, encoder-layers, and decoder-layers are 8, 6, and 6, respectively; the embedding sizes of lyrics, bar\_shift, onset, duration, and onset\_shift are 512, 32, 128, 128, and 128, respectively; dropout is 0.1; batch size is 16; and learning rate is 1e-5 with Adam optimizer. Using a single Tesla-V100-SXM2-32GB GPU to train a satisfactory model takes $\sim$10 hours on the automatically aligned dataset plus 1.5 hours on the manually annotated subset.
In both cases we use 10 percent of the dataset for validation.

\subsection{Rhythm-to-Melody Model}

To train the rhythm-to-melody model, we used POP909 \cite{wang2020pop909} and Lead-Sheet-Dataset \cite{ahu61}.
POP909 contains data on 909 Chinese pop songs including chords, melody in MIDI format, and other information not used here.
Lead-Sheet-Dataset (LSD), a collection of symbolic content sourced from HookTheory,\footnote{\url{https://www.hooktheory.com/}} contains lead-sheets (i.e., melodies and chords) of 16K song segments in MIDI format.
We used the songs in 4/4 time (dropping roughly 10\% of the data)
and transposed all pieces to the key of C major or A minor.
This resulted in about 30K bars of music from POP909 and about 40K from LSD.

We followed \cite{choi2021chord} to train the model, setting the input length to be 8 bars but reducing the pitch range from 48 to 20; any pitches outside the range were octave-shifted to lie within the range. After obtaining the rhythm MIDI of a full song from the lyric-to-rhythm module, pitches were generated for 8 bars autoregressively, with a 4-bar sliding window, i.e., the model composes the next 4 bars given the previous 4 bars already composed.

\section{Evaluation}
\label{sec:exp_res}

We would like to answer two questions: first, does our system succeed in emulating basic musical qualities of the training data? And second, does it produce pleasing, viable settings of lyrics?
To answer the first, we compare the output melodies of our model (denoted `pop-melody') to the held-out training data and discuss their similarity. For the second, we conducted a listening test in which participants rated the quality of the lyric settings of our model as well as those of a state-of-the-art alternative, TeleMelody~\cite{ju2021telemelody}.

\subsection{Objective Results}
\begin{figure}[t!]
\centering
\begin{subfigure}[b]{\columnwidth}
  \centering
  \includegraphics[width=1\columnwidth]{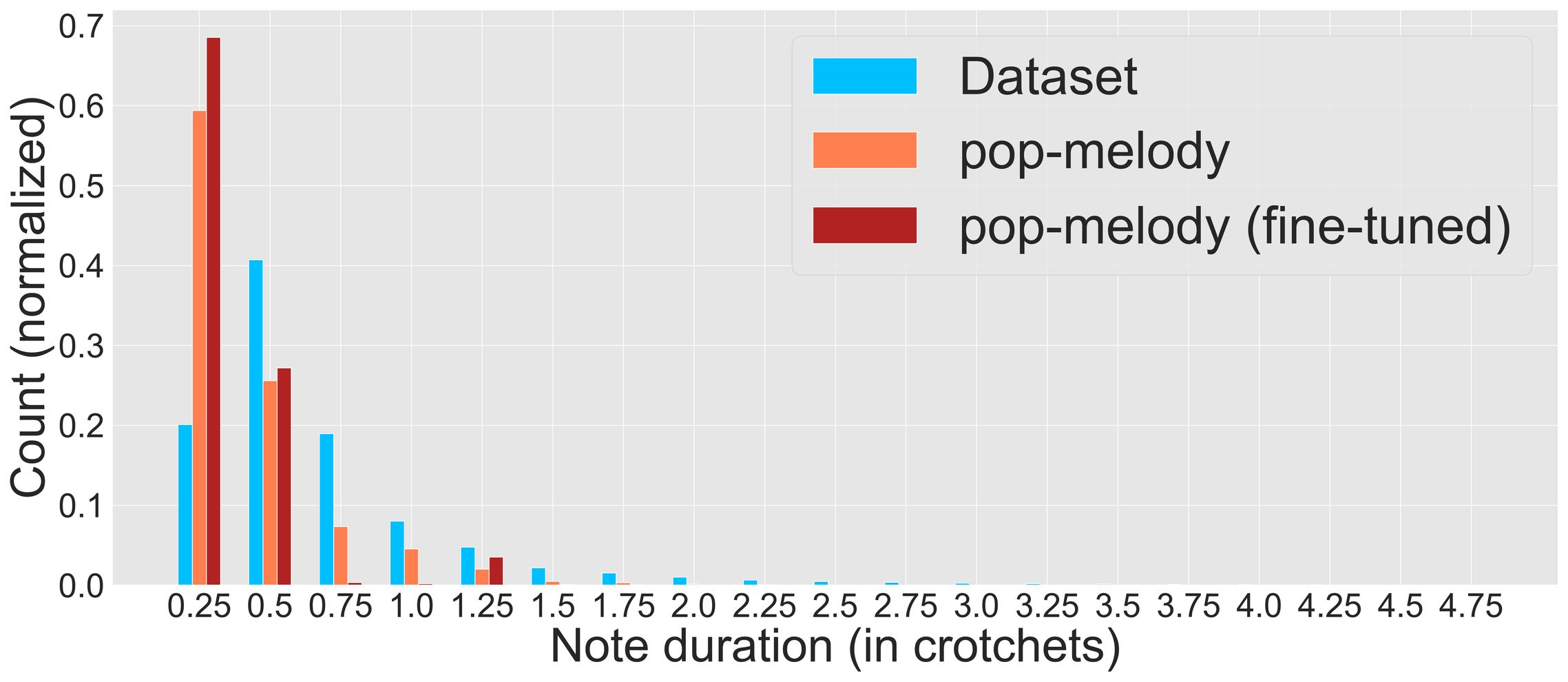}
  \end{subfigure}

    \begin{subfigure}[b]{\columnwidth}
  \centering
  \includegraphics[width=1\columnwidth]{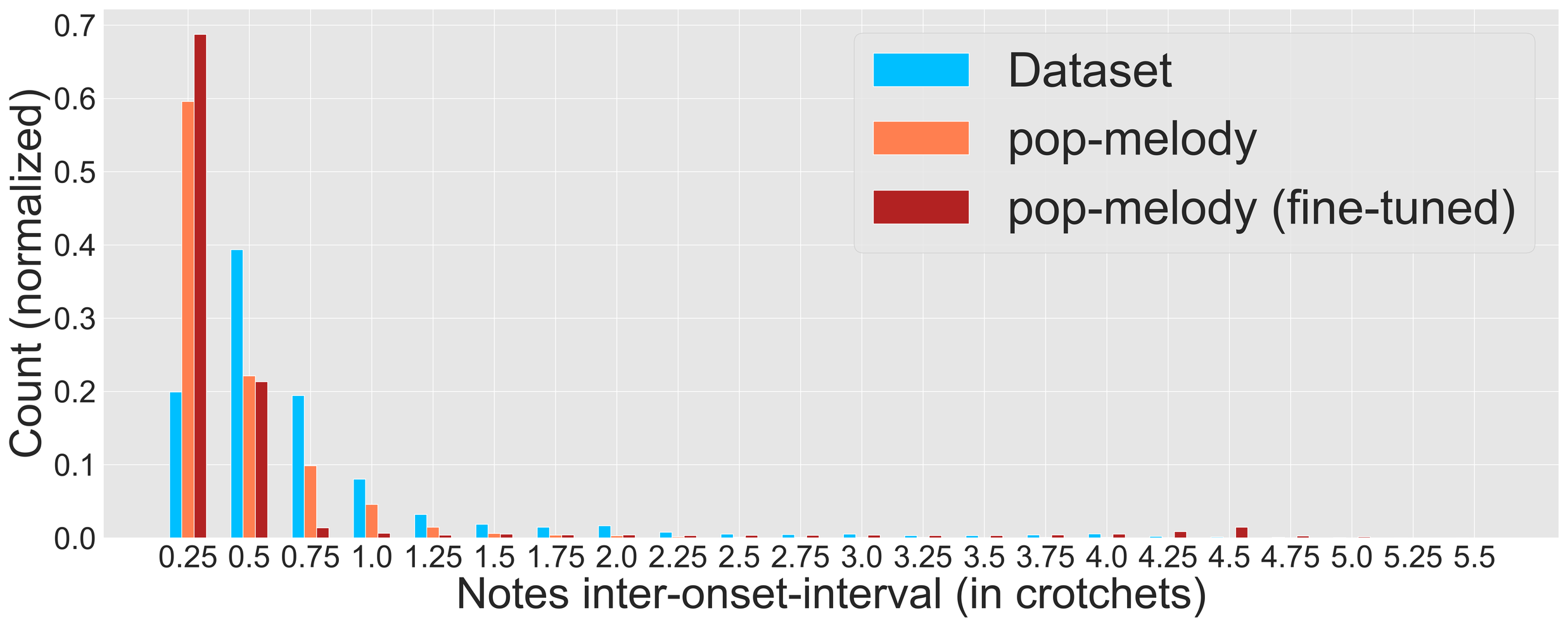}
  \end{subfigure}
  
    \begin{subfigure}[b]{\columnwidth}
  \centering
  \includegraphics[width=1\columnwidth]{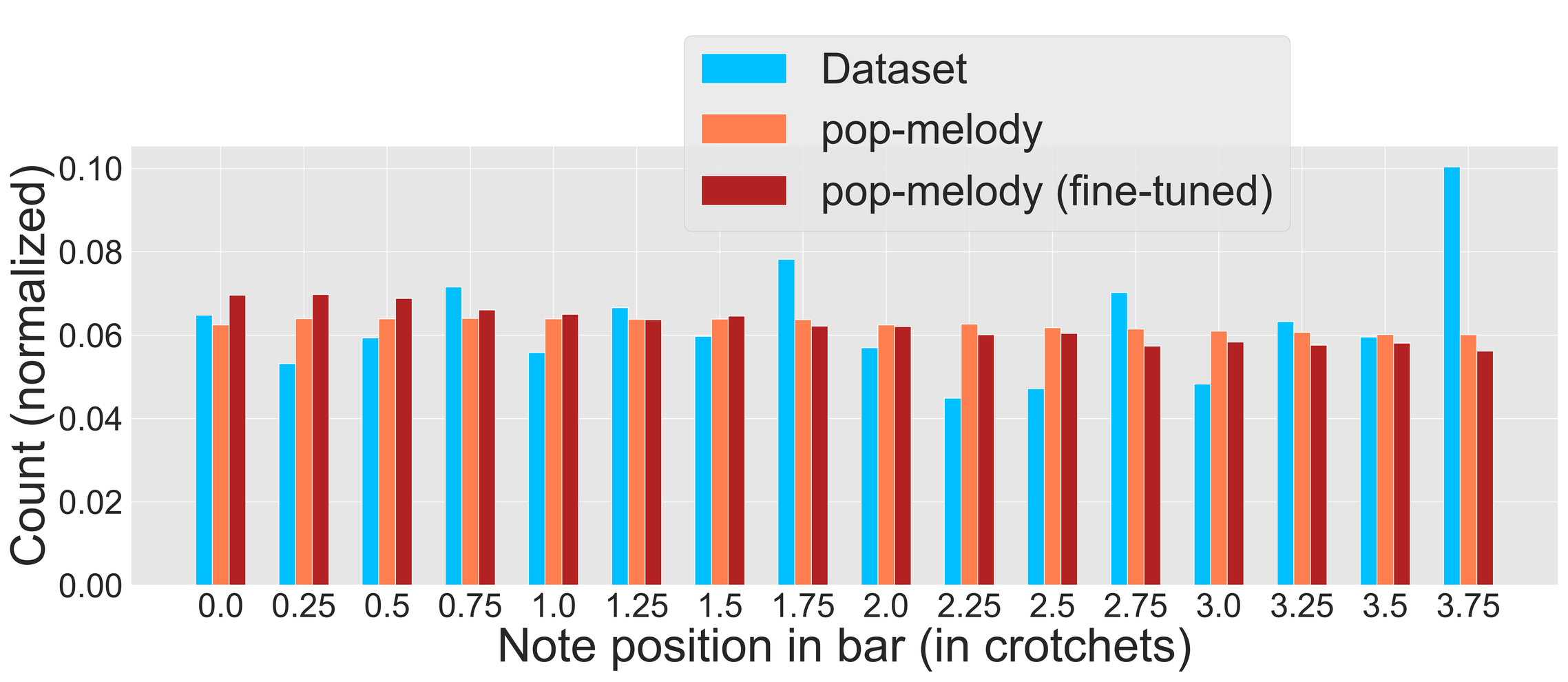}
  \end{subfigure}
 
   \caption{Distributions of descriptors values derived from melodies from the dataset and melodies generated by the proposed pop-melody system.
   }
\label{fig:eval1} 
\end{figure}

\begin{figure}[t!]
\centering
\begin{subfigure}[b]{\columnwidth}
  \centering
  \includegraphics[width=.9\columnwidth]{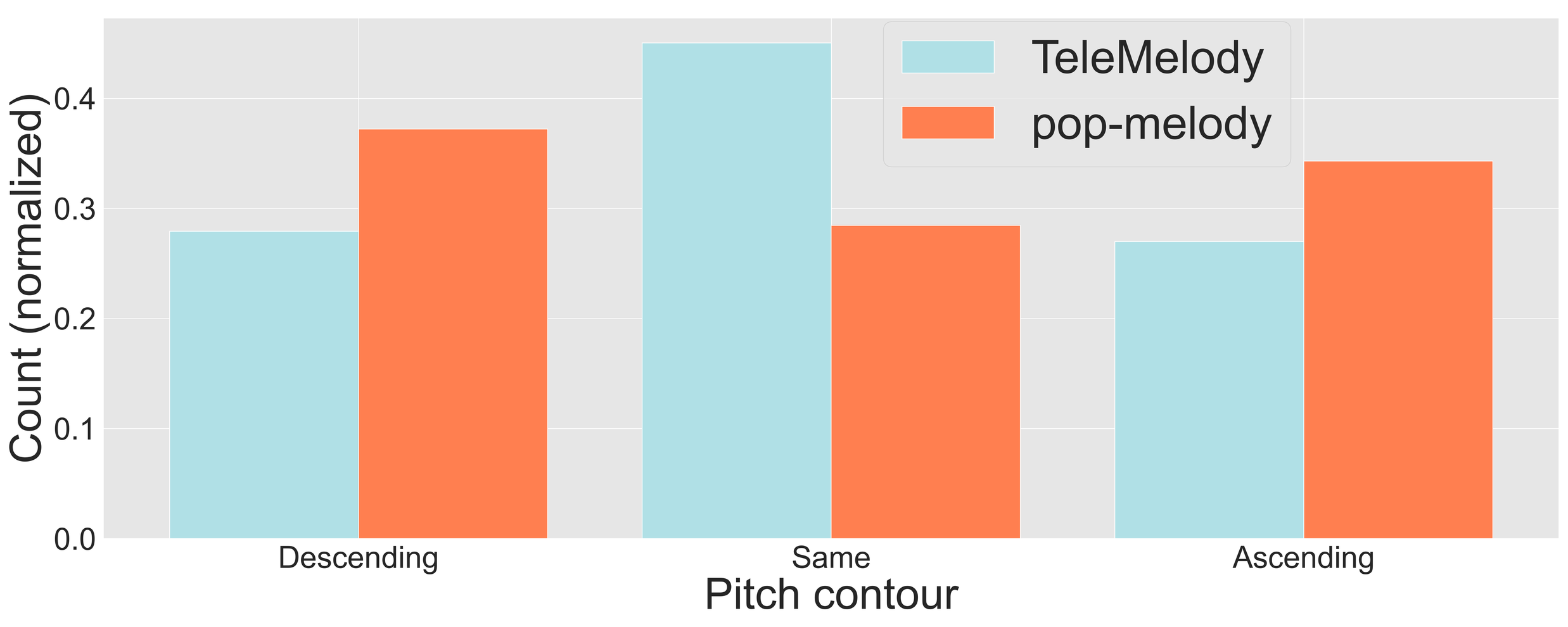}

  \end{subfigure}

    \begin{subfigure}[b]{\columnwidth}
  \centering
  \includegraphics[width=.9\columnwidth]{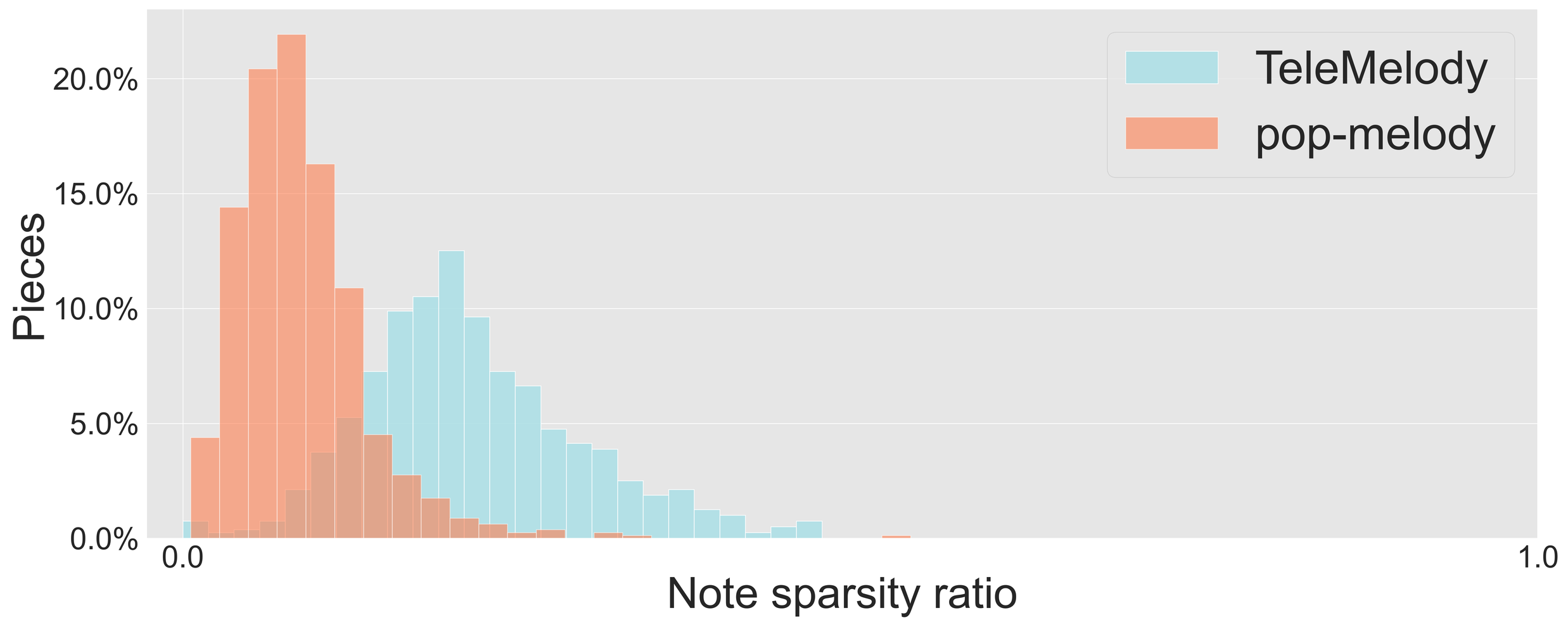}

  \end{subfigure}
  
    \begin{subfigure}[b]{\columnwidth}
  \centering
  \includegraphics[width=.9\columnwidth]{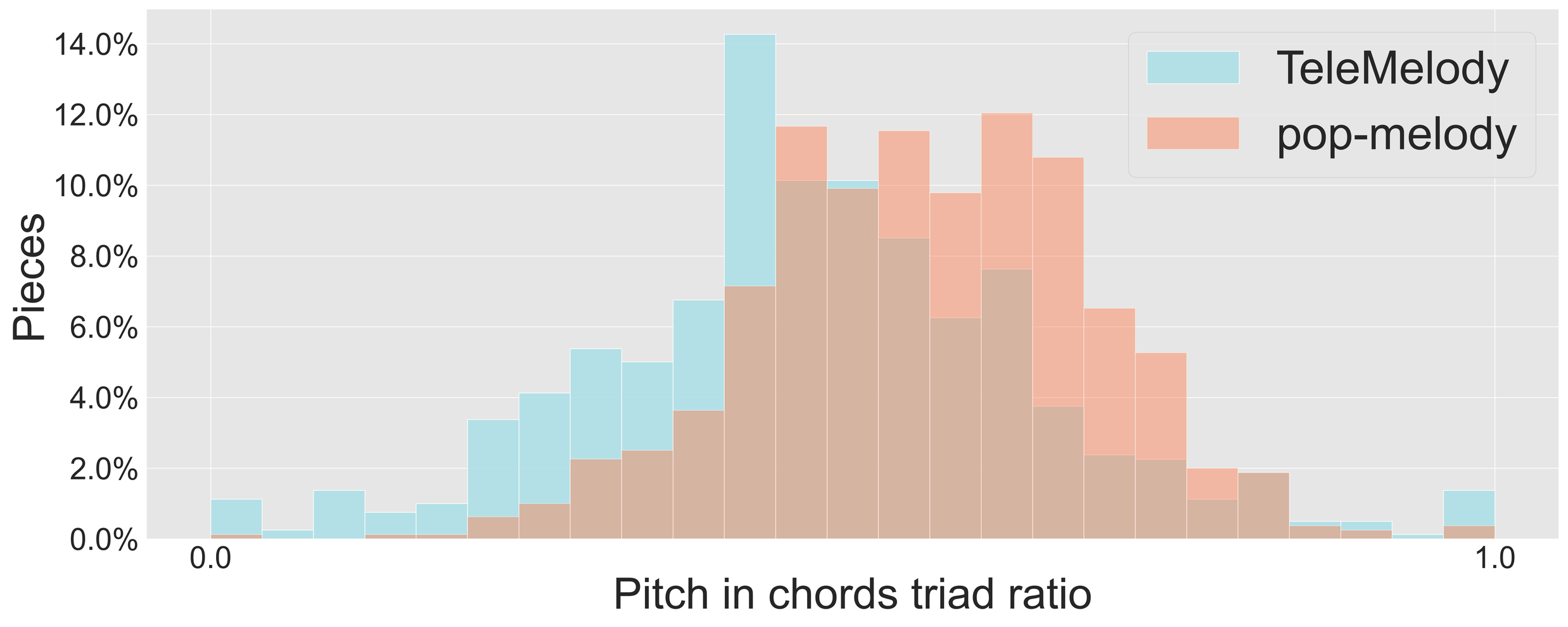}

  \end{subfigure}
    \caption{Distributions of descriptor values derived from melodies generated by TeleMelody~\cite{ju2021telemelody} and by the proposed pop-melody system.}
\label{fig:eval2}
\end{figure}

We analyze the melodies created by our system in two objective evaluation strands.
The first one is to demonstrate how similar the rhythms generated by our model are to the original data (see Fig.~\ref{fig:eval1}); the second is to look for and characterize the differences between the melodies produced by the two models (see Fig.~\ref{fig:eval2}). 
We compare statistics over several musical quantities computed on the dataset and compositions generated by both systems. For this comparison, we have generated 400 scores from each system and used the same amount of scores from the dataset.

Most of the musical quantities we compute are adapted from~\cite{Conklin} and~\cite{pearce2005construction}. These symbolic descriptors have been shown to enhance melodic expectation when embodied in a cognitively plausible system for music prediction.
Expectation and memorability have been shown to be important characteristics for identifying a plausible melody, and surprise and repetition are measurable elements that relate to these charactertistics. (For more background on such descriptors and on the concepts of predictability and uncertainty in the pleasure of music, see~\cite{IDYOM, FANTASTIC}.)

We showcase two sets of descriptors, one for each evaluation strand. The first contains: the \textit{duration} of the melody notes; their \textit{inter-onset intervals} (IOIs; the distance between the start of a note to the start of the preceding one); and their metrical \textit{position in bar}.
Fig.~\ref{fig:eval1} shows the distributions of these descriptors for the dataset and for the outputs of our system (tagged as ``pop-melody'') before and after fine-tuning (see Section~\ref{subsec:l2r}). Note that we exclude TeleMelody from this comparison since it was trained on a different dataset (of around 110K samples), so it is not meaningful to compare it to our training data.

Judging from the distributions, the outputs of both models are broadly similar to the melodies in the dataset.
However, there is clearly a surfeit of short notes (0.25 crochets, or sixteenth notes) in the generated melodies, which skews the distribution of IOIs as a result.
Also, regarding note position in the bar, there is a subtle variation of the likely onset positions in the dataset that is not reflected in the generated data, which, prior to fine-tuning, has an almost uniform distribution.

The other set of descriptors contains: the \textit{pitch contour}, 
which gives the likelihood that the
next note in the melody will be lower (descending), higher (ascending) or the same; \textit{note sparsity}, which gives the fraction of the timeline which has no note in the melody (a value of 0 indicates no rests in the melody); and the \textit{pitch-in-chord-triads ratio}, a kind of `consonance' metric, calculated as the fraction of notes in the melody that belong to the accompanying chord triad.

These descriptors are illustrated in Fig.~\ref{fig:eval2}, comparing the melodies from our fine-tuned system (``pop-melody'') with TeleMelody.
Here, the purpose is not to compare the systems to the data---they were trained on different data, and may each reflect their training set well---but to assess how the melodies of the systems differ.
From the contour descriptors, it is clear that TeleMelody is more likely to generate many consecutive notes with the same pitch, whereas melodies from the proposed system have more variation.
The melodies from our system also tend to have fewer rests, and tend to include more notes that appear in the underlying chord.
The latter can be interpreted as a tendency to stay in consonance and limiting the space of ``dissonant'' or ``unexpected'' moments.




\subsection{Subjective Results}
\label{subsec:subjective_results}
We conducted a subjective listening test using a similar design as~\cite{ju2021telemelody,sheng2021songmass,zhu2018xiaoice}:
we selected lyrics from ten random songs from the test portion of the dataset of 45K songs
and used these as input to three systems to generate melodies: ``TeleMelody''; ``Pop-melody'', the proposed system; and ``Baseline'', an ablated version of our system that does not use POS tokens (see Sec~\ref{subsec:l2r}). This resulted in 30 full songs: ten triples with the same lyrics, chords, and tempo.
We rendered the lyrics and melodies to audio with an in-house singing voice synthesis comparable to Xiaoice~\cite{zhu2018xiaoice} and rendered a simple accompaniment with the chords.

We had 20 participants, all of whom had some musical background and could read musical scores and play an instrument.
Participants listened to a triple at a time, where the system identities were masked and the order is at random. Then, they rated each song on four criteria on a Likert scale from \textit{Bad} (1) to \textit{Excellent} (5):
\begin{enumerate}
    \setlength{\itemsep}{-4pt}%
    \item \textbf{Rhythm}: is the timing of notes suitable for the lyrics? 
    \item \textbf{Harmony}: do the pitches fit the chords and key? 
    \item \textbf{Melody}: does the melody line sound natural with the lyrics? 
    \item \textbf{Overall}: what is the overall quality of the melody?
\end{enumerate}
After rating each triple, listeners also rated their familiarity with the original song of the input lyrics on a 5-point Likert scale. The average rating here was 1.5: somewhere between ``1. Never heard the title or melody'' and ``2. Heard the song title, but not the melody''.

The results of the study are shown in Table~\ref{tab:subjective}.
Overall, listeners gave the three systems similar average ratings: all lie within $3.6\pm0.25$. However, Wilcoxon signed-rank tests reveal small but consistent differences between the systems; see Table~\ref{tab:t-test} for the $p$-values of all comparisons. First, we see that the proposed system is consistently better than Baseline, suggesting that POS-based tokenization is effective. Second, we find that the proposed system also matches or outperforms TeleMelody; the difference is greatest for rhythmic quality.
Despite the broadly positive ratings, mostly between \textit{Fair} (3) and \textit{Good} (4), comments from the participants mostly cited shortcomings of the output. TeleMelody and Pop-melody both earned comments that the ``melody is a little weird'' and sometimes ``too repetitive'', but only the TeleMelody outputs earned comments that the ``rhythm is a little weird'' and ``fragmented''.


%

\begin{table}
\small \centering
 \begin{tabular}{l|cccc}
System & Rhythm & Harmony & Melody & Overall\\
\hline
Baseline  &
3.42(.78) & 3.67(.75) & 3.46(.81) & 3.42(.67) \\
TeleMelody &
3.58(.82) & 3.69(.83) & 3.38(.72) &  3.57(.58)  \\
Pop-melody &
3.84(.72) & 3.87(.69) & 3.64(.70) & 3.68(.57)  \\ 

 \end{tabular}
\caption{Subjective result and comparison.}
\label{tab:subjective}
\end{table}

\begin{table}
\small \centering
 \begin{tabular}{l|cccc}
Comparison & Rhythm & Harmony & Melody & Overall \\
\hline
Pop vs Tele  &    0.0006   &   0.007  &   0.001   &  0.1  \\
Pop vs Baseline  &  1.1e-08  & 0.002  &  0.006  &  1.8e-05   \\
Tele vs Baseline  &  0.01    &  0.89  &  0.22  &  0.02    \\
 \end{tabular}
\caption{P-values of the subjective result comparison.}
\label{tab:t-test}
\end{table}




\begin{figure}[t!]
  \centering
  \includegraphics[width=1\columnwidth]{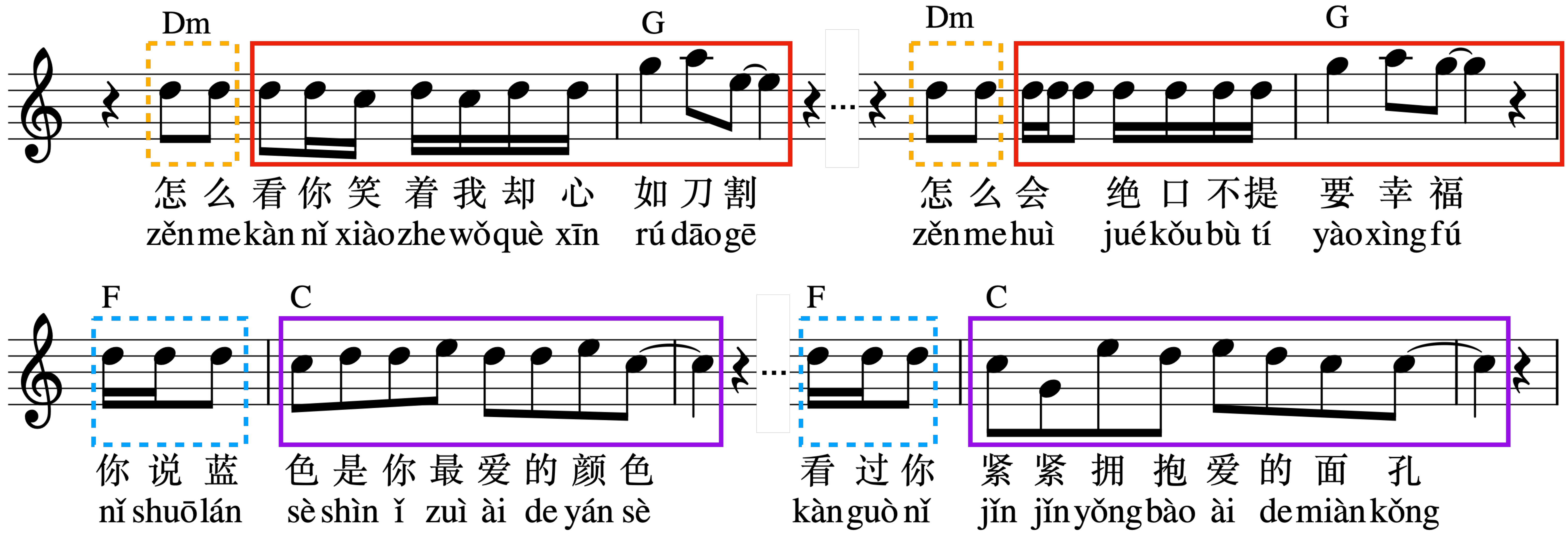}
  \caption{Output examples (a) above and (b) below.}
  \label{fig:out1}
\end{figure}

Two output examples from our system are shown in Figs.~\ref{fig:out1}(a) and~\ref{fig:out1}(b).
In both cases the melodies follow the input chord progressions and we find
parallelism and variation in the melody when the lyric structure recurs.
E.g., in Fig.~\ref{fig:out1}(a), the similar lyrics begin with the same two melody notes (dashed boxes), and the remainders have similar rhythm and contour (solid boxes). Similarly, in Fig.~\ref{fig:out1}(b), the similar lyrics are given identical openings (dashed boxes) with rhythmically identical continuations (solid boxes).

\section{Conclusion and future work}
\label{sec:conclusion}
In this paper we proposed a new approach to generate melody for a given lyric by combining lyric-to-rhythm and rhythm-to-melody modules.
We found that listeners rated the long-term text-settings provided by our system as acceptable, and at least as good as a competing system.

In order to achieve a cross-modal mapping from syllables to onsets to melody notes, we made the simplifying assumption that each syllable is sung on one note.
There is a clear way to improve this in the lyric-to-rhythm module by adding a syllable-state token to the Compound Word, indicating whether we are at the onset of a syllable, or the continuation of one.
However, allowing a one-to-many syllable-to-note mapping would also complicate the automatic syllable alignment step, making the hand-corrected data even more precious.

We also found that POS tags were valuable text tokens; using them led to a boost in text-setting quality. Given this success, we ought to leverage more linguistic information, such as syllable stress and word frequency, as in~\cite{fukayama2010automatic,nichols2009lyric}. 
Music structure labels (e.g., verse and chorus) could also prove valuable. This is an under-explored area, but may become feasible with the introduction of more datasets, or with automatic labeling systems~\cite{wang2022catch} to further augment the existing data.

%




\bibliography{ISMIRtemplate}

\end{document}